\documentclass[aps,prl,twocolumn,showpacs,amsmath,amssymb,superscriptaddress,floatfix]{revtex4-2}
\usepackage{graphicx}
\usepackage{CJK}
\usepackage{color}
\usepackage[colorlinks,bookmarks=false,citecolor=blue,linkcolor=red,urlcolor=blue]{hyperref}
\usepackage{multirow}
\usepackage{ulem}
\usepackage{tabularx}
\usepackage{diagbox}
\usepackage[nounderscore]{syntax}
\graphicspath{{fig/}}

\newcommand{\be}{\begin{equation}}
\newcommand{\ee}{\end{equation}}

\newcommand{\trace}{{\rm Tr}}

\begin{document}
	\title{Diagnosing $SO(5)$ Symmetry and First-Order Transition in the $J-Q_3$ Model via Entanglement Entropy}
\author{Zehui Deng}
\altaffiliation{The two authors contributed equally to this work.} 
\affiliation{Beijing Computational Science Research Center, Beijing 100193, China}
\author{Lu Liu}
\altaffiliation{The two authors contributed equally to this work.} 
\affiliation{School of Physics, Beijing Institute of Technology, Beijing 100081, China}
\author{Wenan Guo}
\email{waguo@bnu.edu.cn}
\affiliation{Department of Physics, Beijing Normal University, Beijing 100875, China}
\affiliation{Key Laboratory of Multiscale Spin Physics (Ministry of Education), Beijing Normal University, Beijing 100875, China}
\affiliation{Beijing Computational Science Research Center, Beijing 100193, China}
\author{Hai-Qing Lin}
\email{hqlin@zju.edu.cn}
\affiliation{Institute for Advanced Study in Physics and School of Physics, Zhejiang University, Hangzhou 310058, China}
\date{\today}

\begin{abstract}
We study the scaling behavior of the R\'enyi entanglement entropy with smooth boundaries at the phase 
transition point of the two-dimensional $J-Q_3$ model. 
Using  the recently developed scaling formula [Deng {\it et al.}, Phys. Rev. B {\textbf{108}, 125144 (2023)}], 
we find a subleading logarithmic term with a coefficient showing that the number of Goldstone modes is four, 
indicating the existence of the spontaneous symmetry breaking from an emergent $SO(5)$ to $O(4)$ in the thermodynamic 
limit, but restored in a finite size.
This result shows that the believed deconfined quantum critical point of the $J-Q_{3}$ model is 
a weak first-order transition point. 
Our work provides a new way to distinguish a state with spontaneously broken continuous symmetry from 
a critical state. The method is particularly useful in identifying weak first-order phase transitions, which are hard to determine using conventional methods. 
\end{abstract}

\maketitle

{\it Introduction.}---Deconfined quantum criticality (DQC), which describes continuous phase transition 
between two unrelated ordered states, is beyond the paradigm of Landau-Ginzburg-Wilson\cite{Senthil, senthil_PRB}. 
Sandvik invents the $J-Q_2$ model\cite{Sandvik2007}, which realizes the valence-bond solid (VBS)-N\'eel transition in 
two-dimensional (2D) quantum spin systems. The 
model has no sign problem and, therefore, is amenable to quantum Monte Carlo (QMC) simulations to study DQC.  
The $J-Q_3$ model \cite{JieLou} is a variant of the $J-Q_2$ model showing a similar VBS-N\'eel transition 
but with the VBS order enhanced in the VBS phase.
Lots of QMC studies of these and other variants of the $J-Q$ model, as well as three-dimensional classical loop models and fermionic models, have characterized the signatures of the DQC\cite{MelkoPRL2008, Kaul_su3, kawashima_SUN, 
damle_honeycomb, shao_science, Nahum_PRX2015, Liuyh2019, Yinshuai2022}, and shown that the observed quantum phase transition appears to be continuous. 

The N\'eel-VBS DQC can also be described by a nonlinear sigma model containing a Wess-Zumino-Witten term for the five-component superspin\cite{Senthil_NLSM}. 
The leading anisotropy plays the role of the mass term in the field theory, which drives the transition between the N\'eel and VBS phases. 
Nahum {\it et al.} \cite{Nahum_SO5} conjecture that all the higher anisotropies are irrelevant in the Renormalization Group sense, and there is an 
emergent $SO(5)$ symmetry at the deconfined quantum critical point (DQCP). 
	The conjecture was verified numerically at the DQCP of the loop model\cite{Nahum_SO5}.
The $SO(5)$ symmetry has also been shown explicitly at the VBS-N\'eel transition point of the $J-Q_6$ model \cite{Takahashi_SO5}. Unfortunately, the 
transition is shown to be strongly first-order in sharp contrast to the $J-Q_2$ and $J-Q_3$ models.

The conformal bootstrap calculation based on $SO(5)$ symmetry sets bounds on  the correlation-length exponent and the anomalous dimension of a critical point\cite{nakayama2016}. The exponents of the ``continuous'' VBS-N\'eel transitions \cite{shao_science, Sandvik2007, Nahum_PRX2015, sandvik_cpl} do not satisfy 
the bounds, alternative scenarios are suggested:
The transition is described by a nonunitary conformal field theory (CFT) with complex fixed points 
\cite{Nahum_quasiuniv, MaRC_pseudocri}; a multicriticality is involed\cite{Zhaobw_multicri, Ludc_multicri}; or
it is precursors to a weak first-order transition\cite{Kuklov_firstorder, chenk_firstorder, Jiang_2008}.

Entanglement entropy (EE) is a valuable tool to resolve the puzzle.   
At the criticality of a (2+1)D system, besides the area law, the scaling of EE has a logarithmic 
term with a {\it negative} coefficient when the boundary has sharp corners but no logarithmic term
if the boundary is smooth\cite{c-theorem}.  
The scaling behavior of the R\'enyi EE of the $J-Q_3$ model at the transition point 
has been studied recently \cite{ZhaoJR_JQ3}: 
a positive corner logarithmic term is found, in sharp contradiction with the prediction of unitary CFT.
However, a very recent paper shows that the positive logarithmic term due to corners becomes negative when the 
tilted bipartitioning is applied \cite{Jon2024}.

It is worth noting that if the (2+1)D system is ordered with continuous symmetry broken, the scaling of EE 
also has a logarithmic 
term with a {\it negative} coefficient when the boundary has sharp corners\cite{casini2007universal}. In addition, 
another logarithmic term with a coefficient proportional to the number of Goldstone modes in the scaling form 
of EE is present even when the boundary is smooth\cite{Metlitski-EE}
\begin{equation}
	S_n(L) =a L^{d-1}+\frac{N_G}{2} \ln(\frac{\rho_{s}}{c}L^{d-1})+\gamma_{\rm ord},
    \label{fssS2}
\end{equation}
where $L$ is the system size, $d$ is the spatial dimension, $N_G$ is the number of Goldstone modes of the ordered phase, $\rho_s$ is the spin stiffness, and $c$ is 
the spinwave velocity. $\gamma_{\rm ord}$ is a universal geometry-dependent finite constant, as all the 
short-distance physics are absorbed into $\rho_s$ and $c$.
Unfortunately, this formula only works at very large system
sizes\cite{D'Emidio,zhao2022measuring} or the order is very strongly enhanced\cite{D'Emidio}, or the 
continuous symmetry is $O(2)$\cite{Kulchytskyy}. 

In a recent work \cite{Deng_PRB}, we propose a modified scaling formula for the EE with smooth boundary
 \begin{equation}
        S_{n}(L) =a L^{d-1} + \frac{N_G}{2} \ln(I(L)^{1/2}\rho_s(L)^{1/2} L^{d-1}) +\gamma_{\rm ord},
        \label{newS2}
\end{equation}
where $I(L)$ is the finite size inertia moment density of the quantum rotor describing the energy spectrum of the 
$O(n)$ ordered model. Using this formula, 
The correct $N_G/2$ is extracted from data of rather small 
system sizes for the 2D Heisenberg model and the bilayer Heisenberg model.  
The reason why Eq.(\ref{fssS2}) works for $O(2)$ symmetry is also explained by pointing out the 
leading order correction in $I(L)$ vanishes\cite{Deng_PRB}. 

\begin{figure}[h]
    \centering
    \includegraphics[width=0.45\textwidth]{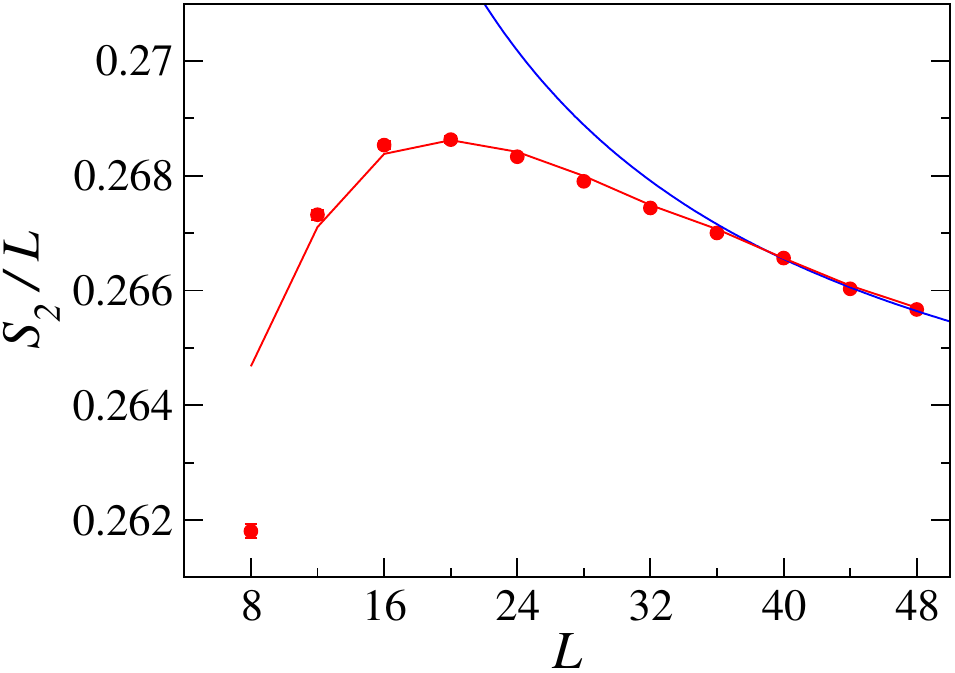}
    \caption{$S_2(L)/L$ versus system size $L$ at $Q/J=1.49153$.  
    The red solid line is the fit of Eq. (\ref{newS2}) for $L\geq L_{min}=24$ with $N_G/2$ found to be $2.01(14)$, see Tab.\ref{tab:fitlog_mod} for other parameters. 
  The blue solid line shows the fit without logarithmic corrections for $L\geq L_{min}=40$, see Tab. \ref{tab:arealaw}.
    }
    \label{fig:S2-L}
\end{figure}

In this letter, we calculate the R\'enyi EE $S_2(L)$ with {\it smooth} boundaries at the 
transition point of the $J-Q_3$ model; the results are shown in Fig. \ref{fig:S2-L}. Our data shows 
the presence of a logarithmic term in the scaling form, which has also been reported very 
recently\cite{songmh2023}, but not found when tilted bipartitioning is applied\cite{Jon2024}.
We then use the modified formula Eq. (\ref{newS2}) to show that, at the believed DQCP of the $J-Q_3$ model, 
the system bears an emergent $SO(5)$ symmetry, which is broken spontaneously but restored in a finite size. 
We first calculate $\rho_s(L)$, then
determine the inertia moment density $I(L)$  by calculating magnetization as a function of the magnetic field.
These are done via standard QMC simulations.       
Using $I(L)$ and $\rho_s(L)$ as inputs, we fit Eq. (\ref{newS2}) to the obtained $S_2(L)$.  We 
find $N_G=4$ at the transition point, showing the finite-size system has the $SO(5)$ symmetry, 
described by the vector formed by the $O(3)$ N\'eel order and $O(2)$ VBS order,
breaking into an $O(4)$ symmetry in the thermodynamic limit. 
Since the formula applies only to the ordered phase,
this also proves that the transition is first order, similar to what happens in the checker-board 
$J-Q$ model \cite{zhaobw_nphysics}. 

\begin{figure}
    \centering
    \includegraphics[width=0.3\textwidth]{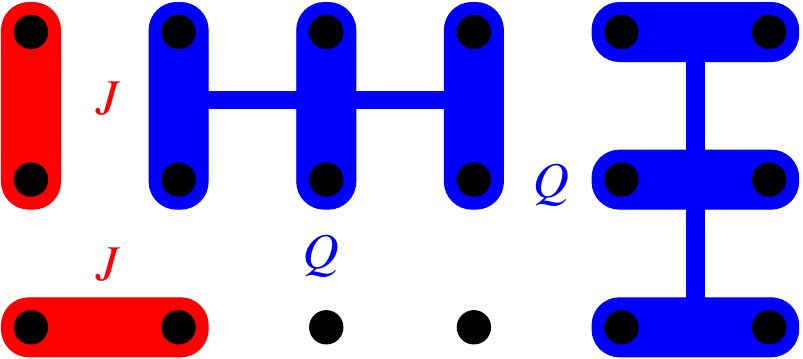}
	\caption{Illustration of $J$ terms (light red bars) and $Q$ terms (three connected blue bars) in the 2D $J-Q_3$ model. }
    \label{fig:jq3model}
\end{figure}

	\textit{Model and method.}---
The $J-Q_3$ model on a 2D square lattice is described by the following Hamiltonian
\begin{equation}
	H= -J\sum_{\langle i, j\rangle} C_{ij}  - Q\sum_{\langle ijklmn\rangle}C_{ij}C_{kl}C_{mn}-h \sum_i S_i^z,
\end{equation}
where $C_{ij}=\frac{1}{4}-{\bf S}_{i}\cdot {\bf S}_{j}$ is the singlet projector at sites $i$ and $j$. The nearest-neighbor $J$ terms and the three parallel projector product $Q$ terms are illustrated in Fig. \ref{fig:jq3model}. 
$h$ is an external magnetic field. 
Without the external magnetic field, a phase transition separates the N\'eel ground state for small $Q/J$ and the VBS state for large $Q/J$. The latest estimate of the transition point is $Q/J=1.49153(31)$
\cite{WangYC_DisorderOP}.

The R\'enyi EE is defined as
\begin{equation}
    S_{n}(A)=\frac{1}{1-n}\ln{\trace [\rho_{A}^{n}]},
    \label{Sn_def}
\end{equation}
where  $\rho_{A}=\trace_{\bar{A}}{\rho}$ is the reduced density matrix of a 
subsystem $A$ with $\bar{A}$ its complement, $n$ is the R\'enyi index( $n=2$ in our work). 
$\rho=e^{-\beta H}/Z$ is the density operator with $Z=\trace e^{-\beta H}$ the partition function. $\beta \to \infty$ 
is the inverse temperature to probe only the properties of the ground states.

In the QMC simulations, we consider an $L\times L$ square lattice with periodic boundary conditions 
employed in both lattice directions. In particular, to calculate $S_{2}(A)$, we consider bipartite the toroidal 
lattice into two equally sized cylindrical strips of size 
$N_{A}=L/2\times L=N_{\bar{A}}$ containing no corners and study the R\'enyi EE of one subregion. 

We use the nonequilibrium work algorithm developed recently \cite{D'Emidio} in the version of the 
projector quantum Monte Carlo (PQMC) method \cite{sandvik-pqmc,pqmc-loopupdate} to extract the R\'enyi EE. 
In the specific simulations, to guarantee the accuracy of the EE, we compute 2000 to 3000 nonequilibrium 
work realizations for each system size ranging from $L=8$ to $L=48$. Each work realization consists of 
$N_{A}\times 10,000$ nonequilibrium times steps. The projection power $m=L^3$ in the PQMC simulations.

The spin stiffness $\rho_s$ is defined as the free energy increasing due to the presence of a twist field.
We apply the stochastic series expansion (SSE) QMC  simulation with the loop update 
algorithm at inverse temperature $\beta=2L$\cite{Sandvik1997, Sandvik-review} to calculate $\rho_s$ 
through the fluctuations of the winding number of spin transporting
\begin{equation}
    \rho_{s}=\frac{n}{4\beta N}\langle{ L_x^2 W^{2}_{x}+ L_y^2 W^{2}_{y}}\rangle,
    \label{stiffness}
\end{equation}
where the winding numbers are defined as
\begin{equation}
    W_{\alpha}=(N^{+}_{\alpha}-N^{-}_{\alpha})/L_\alpha.
\end{equation}
Here, $N^{+}_{\alpha}(N^{-}_{\alpha})$ is the total number of operators transporting spin in the positive (negative) $\alpha=x, y$ direction. 
$N=L_x\times L_y$ is the number of spins. The factor $n$ is included to 
account for rotational averaging for the systems with $O(n)$ symmetry. We have set $n$ to be the tempted value
5. The choice of $n$ could change neither the presence of the logarithmic term nor the coefficient of the 
term, except for $\gamma_{\rm ord}$. Thus, our choice will not bias our conclusion and will show the 
correct self consistently.   

 \begin{figure}[h]
     \centering
     \includegraphics[width=0.45\textwidth]{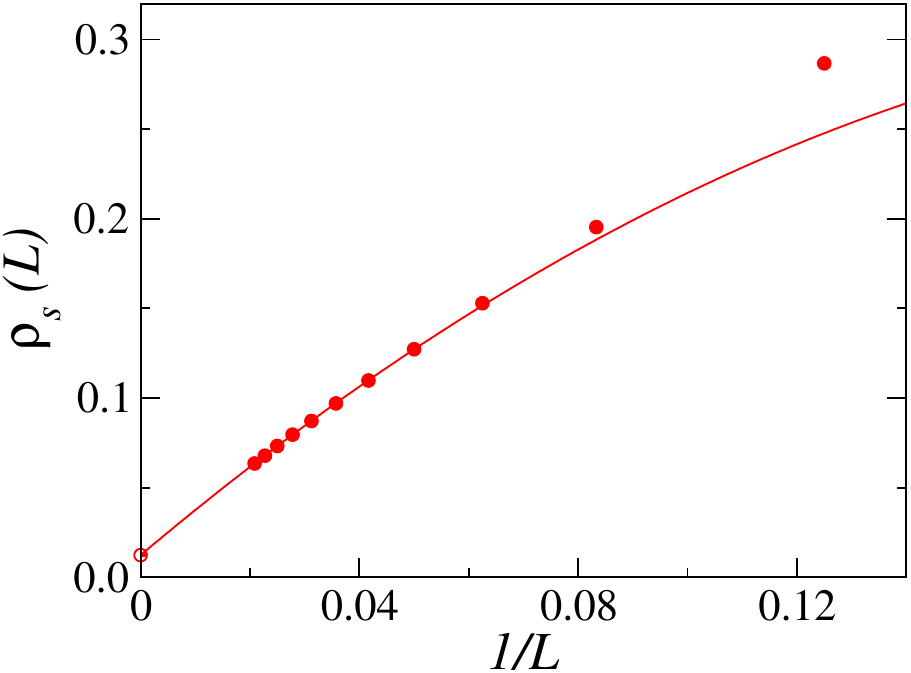}
     \caption{ Spin stiffness $\rho_s(L)$ versus $1/L$ at $Q/J=1.49153$. The red solid line is a polynomial fit for data points with $L\ge 24$.}
     \label{chirhos_H}
 \end{figure}

 In Fig. \ref{chirhos_H}, we show the spin-stiffness $\rho_{s}$ for each system size at $Q/J=1.49153$. Polynomial fitting shows $\rho_s(L)$ converges to a finite value.
 However, power-law fitting 
 with an exponent $\sim -0.78$ is also statistically sound. See Ref.\cite{supplemental} for details.
 This is similar to what is found at the transition point of the $J-Q_2$ model \cite{Jiang_2008}.
For conventional $(2+1)$D critical behavior, $\rho_s(L)$ should scale as $1/L$.
One interpretation of this unusual scaling behavior is that the transition is first order,
it can also be accounted for by an unconventional two-length scales scaling scheme\cite{shao_science}.
In this work, we do not try to determine or explain the finite-size scaling behavior of $\rho_s(L)$ and $I(L)$. Instead,
we use the finite-size value of $\rho_s(L)$ and $I(L)$ as inputs of the fitting formula Eq. (\ref{newS2}).
The results of the fits, in turn, support $I(L)$ and $\rho_s(L)$ have finite values at the thermodynamic limit.


To make use of Eq. (\ref{newS2}), we need to determine the inertial density $I(L)$ as well. 

Suppose the model is described by $O(N)$ quantum rotors with $S$ being the total superspin of the 
system.
The energy of the tower of excited states is 
\begin{equation}
	E_L(S)=\frac{S (S+N-2)}{2L^2 I(L)},
 \label{energylevels1}
\end{equation}
where $I(L)$ is the inertia moment density and $L$ is linear size of the system. 
Here, we have set $E_L(0)=0$. 

At the thermodynamic limit, $I(L)$ converges to the transverse susceptibility $\chi_\perp$.  
The chiral perturbation theory \cite{chiral-perturbation} predicts the finite-size 
behavior of $I(L)$ up to $1/L$ \cite{Neuberger_fss_AFH, FisherDS_fss_AF, PhysRevLett.80.1746}. 
However, for the current model, we find the chiral perturbation theory does not work due to the
extremely small 
value of $\chi_\perp$, resulting in that the order $1/L$ is insufficient, see Ref. \cite{supplemental}.
We then try to find out the energy levels $E_L(S)$ and calculate $I(L)$ from Eq. (\ref{energylevels1}) directly.  

This is done by adding a magnetic field and studying the field dependence of the magnetization
$M_z=\langle m_z \rangle$,
with $m_z=\sum_i S_i^z$, via the SSE QMC simulation with the directed loop algorithm \cite{directed_loops}.
When a magnetic field is applied to the system along the direction of one component of the superspin, e.g., $S_z$, the energy levels become
\begin{equation}
	E_L(S, m_z)= E_L(S) -h m_z. 
 \label{energylevel}
\end{equation}
At a given inverse temperature $\beta$, the total magnetization as a function of $h$ can be calculated using 
\begin{equation}
M_z (L, h) =\frac{1}{Z}\sum\limits_{S=0}^{L^2/2}\sum\limits_{m_z=-S}^S m_z g(S,m_z) e^{-\beta E_L(S,m_z)},
\label{Mzh-SO}
\end{equation}
with partition function
\begin{equation}
    Z={\sum\limits_{S=0}^{L^2/2}\sum\limits_{m_z=-S}^S g(S,m_z) e^{-\beta E_L(S, m_z)}},
\end{equation}
where $g(S,m_z)$ is the degeneracy for fixed $S$ and $m_z$. This degeneracy changes with the symmetry 
\cite{1959angular}. For example, $g(S, m_z)=1$ in $SO(3)$, however, in $SO(5)$, $g(0,0)=1$, $g(1,0)=3$, $g(1,\pm1)=1$, $g(2,0)=6$, $g(2,\pm1)=3$, $g(2,\pm2)=1$, $\cdots$ \cite{sczhangso(5)}.
Fitting Eq. (\ref{Mzh-SO}) to numerically obtained $M_z(L, h)$ curve, we can obtain $E_L(S)$ and $I(L)$ followed.

\begin{figure}
    \centering
    \includegraphics[width=0.45\textwidth]{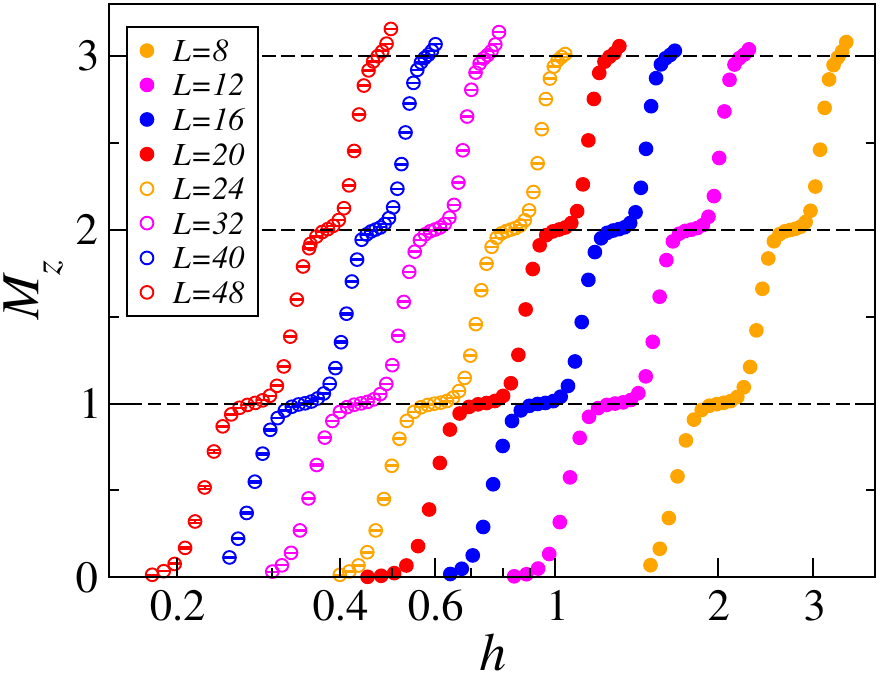}
    \caption{Total magnetization $M_z$ versus external magnetic field $h$ at $Q/J=1.49153$ for different system size $L$.}
    \label{fig:mzvsh}
\end{figure}

\begin{figure}
    \centering
    \includegraphics[width=0.45\textwidth]{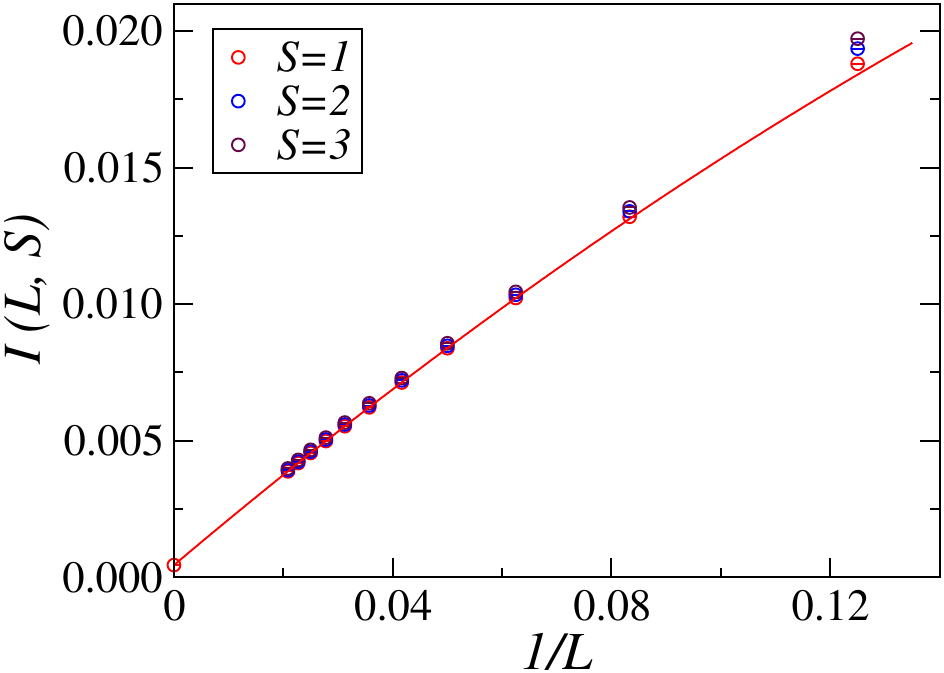}
    \caption{$I(L,S)$ of the $J-Q_3$ model at $Q/J=1.49153$. When $L$ becomes large, $I(L, S)$ converges to the same value for different $S$, showing the emergence of $SO(5)$ symmetry.
	The value $I(L=\infty, S=1)$ is found to be 0.00045(2) using a polynomial fit to data points with size $L\geq20$ (the red solid line), see \cite{supplemental} for details.}
    \label{fig:ILqc}
\end{figure}

Figure \ref{fig:mzvsh} shows $M_z$ as functions of $h$ for different system sizes $L$ at $Q/J=1.49153$.
In the simulations, the inverse temperature is set to $\beta=2L$ to match the lowest rotor levels. 
We thus find $E_L(S)$ and $I(L, S)$, which depends on $S$ at finite size, similar to that in the 2D Heisenberg model \cite{PhysRevLett.80.1746,directed_loops}. 
However, $I(L, S)$ for different $S$ converges to $I(L)$  for large enough $L$, as illustrated in Fig. \ref{fig:ILqc}.
It is then reasonable to use $I(L, S=1)$ as $I(L)$ when fitting the EE $S_2(L)$ according to the scaling formula Eq.(\ref{newS2}). 

\begin{figure}[h]
    \centering
     \includegraphics[width=8cm]{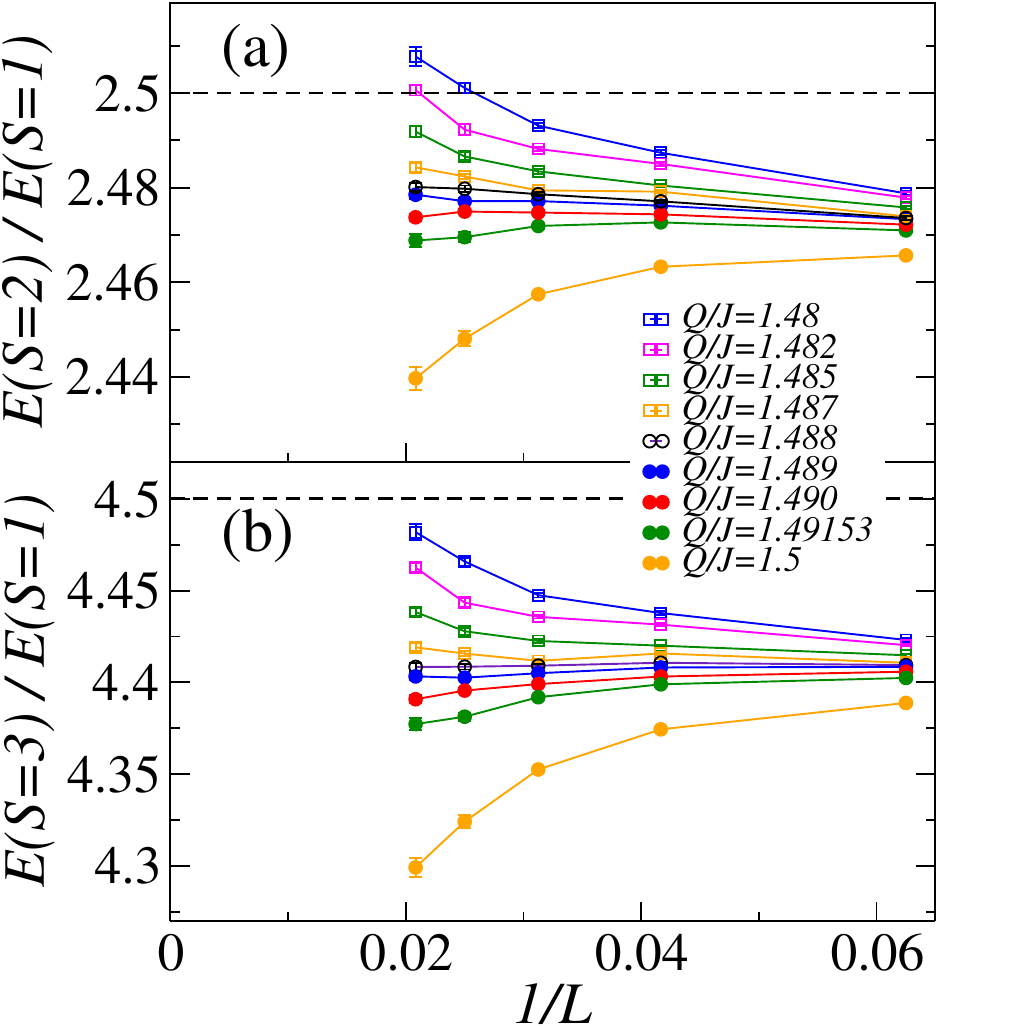}
    \caption{  Energy level ratio versus $1/L$ for $Q/J$ near the transition point. (a) $E(S=2)/E(S=1)$ vs $1/L$; (b) $E(S=3)/E(S=1)$ vs. $1/L$.  
	The dash lines indicate ratios of $SO(5)$ rotor at thermodynamic limit.}
    \label{fig:energy_ratio_zoom}
\end{figure}

As a by-product, we find that $E_L(S=2)/E_L(S=1)$ tends to go to 2.5 and $E(S=3)/E(S=1)$ tends  
to 4.5  for $Q/J$ being around 1.489, slightly larger than the estimated transition point $Q/J=1.49153(31)$ as 
$L$ increases, as illustrated in Fig. \ref{fig:energy_ratio_zoom}.
In the case that the transition point has an emergent $O(5)$ symmetry, 
these behaviors indicate the validity of the excitation spectrum Eq. (\ref{energylevels1}) for small $S$: 
as $L \to \infty$, $I(L, S)$ converge to $I(L=\infty)$, 
the ratios converge to $2.5$ and $4.5$, respectively. Meanwhile, it also suggests that $Q/J=1.489$ could be 
the true transition point. 
We have also done calculations of $S_2(L)$, $\rho_s(L)$, and $I(L, S)$ at $Q/J=1.489$. The results are presented in supplemental materials \cite{supplemental}.  

\textit{Scaling behavior of $S_2(L)$ at transition point.}---
Suppose the system sits at an ordinary (2+1)D critical point. We expect there to be no logarithm correction to 
the area law of $S_2(L)$ since the boundaries separating the subsystem and its complement are smooth. 
The curve of $S_2(L)$ in Fig. \ref{fig:S2-L} should be fitted using 
$S_2(L)/L=a +c/L.$ Table \ref{tab:arealaw} shows the results of such fitting.  The fits are 
statistically sound only for the largest three sizes. 
\begin{table}[thb]
  \caption{ Fitting results of $S_2(L)=aL+c$ to $S_2(L)$ data at $Q/J=1.49153$ with $L=8-48$. 
  }
    \begin{tabular}{c|c|c|c}
     \hline
     \hline
        $L_{min}$ &  $a$           &  $c$                      &$\chi_r^2$/P-value \\ 
     \hline
                36&  0.2616(2)     &   0.194(6)                & 3.35/0.035   \\  
     \hline
                40&  0.2611(3)     &   0.218(12)               & 1.32/0.25   \\
     \hline
    \hline
     \end{tabular}
     \label{tab:arealaw}
 \end{table}

On the other hand, we try to fit the data using the following formula with a logarithmic correction term
\begin{equation}
	S_2(L)/L=a+b\ln(L)/L+c/L,
	\label{S2log}
\end{equation}
which is essentially Eq. (\ref{fssS2}) with $b=N_G/2.$
Table \ref{tab:fitlog} shows details of the fitting procedure.  
The formula can be fitted for all $L \geq 12$ finite-size data.
The presence of a logarithmic correction to the area law is evident. 
However, the thus found $b$ is far away from $N_G/2=1$ of the AF ordered phase, nor $N_G/2=2$ of the expected 
$SO(5)$ ordered state.

\begin{table}[thb]
  \caption{ Fitting results of Eq. (\ref{S2log}) to $S_2(L)$ at $Q/J=1.49153$ with $L=8-48$.}
    \begin{tabular}{c|c|c|c|c}
    \hline
     \hline
        $L_{min}$ &  $a$         & $b$        & $c$                       &$\chi_r^2$/P-value \\   
    \hline
                8 &  0.2569(1)   & 0.215(3)   &  -0.409(5)                & 2.18/0.03   \\
    \hline
               12 &  0.2567(2)   & 0.222(4)   &  -0.424(10)               & 1.99/0.05\\
    \hline
               16 &  0.2563(3)   & 0.233(7)   &  -0.452(17)               & 1.69/0.12 \\
    \hline
               20 &  0.2558(4)   & 0.248(12)  &  -0.487(30)               & 1.59/0.16\\     
    \hline
               24 &  0.2553(6)   & 0.267(19)  &  -0.536(46)               & 1.52/0.19\\
     \hline
    \hline
     \end{tabular}
     \label{tab:fitlog}
 \end{table}

This phenomenon is understandable from the former experience with the 2D AF Heisenberg and the bilayer Heisenberg models. 
The formula Eq.~(\ref{fssS2}) only works at very large systems, or the order is strongly enhanced\cite{D'Emidio}. To see the expected logarithmic correction due to Goldstone modes, we have to 
make use of our improved scaling formula Eq. (\ref{newS2}) to the fitting of the finite-size $S_2(L)$ data. 
The details of the fitting are shown in Table \ref{tab:fitlog_mod}. 
The fits are statistically sound for all $L \geq 12$ finite-size data. 
The results of $N_G/2$ approach 2 within about one error bar for $L_{min}=20$.  The fits remain 
stable upon further excluding small size points by gradually increasing $L_{min}$,
although error bars on the fit parameters increase rapidly. 
Thus, we may safely conclude that the system is ordered with $N_G=4$. 

We did similar scaling analyses for $Q/J=1.489$ and obtained similar results, suggesting the scaling behavior is 
robust in the neighborhood of the transition point. See Ref. \cite{supplemental} for details.  

 \begin{table}[thb]
  \caption{Fitting results of Eq. (\ref{newS2}) to $S_2(L)$ at $Q/J=1.49153$ with $L=8-48$. In the fits, finite-size $\rho_s(L)$ and $I(L,S=1)$ are used as 
	 inputs. For results obtained using $I(L, S=2)$ and $I(L, S=3)$, see \cite{supplemental}.
  }
    \begin{tabular}{c|c|c|c|c}
    \hline
     \hline
         $L_{min}$ &  $a$    & $b$& $\gamma_{ord}$&$\chi_r^2$/P-value \\   
    \hline
                12 &  0.2540(2)  &  1.89(4) &   1.10(2)                & 0.93/0.48   \\
    \hline
                16 & 0.2543(3)   & 1.83(6)   & 1.07(3)                 &0.82/0.55\\
   \hline
               20  &  0.2541(5)  & 1.88(10)   &  1.09(5)                & 0.91/0.47 \\     
    \hline
               24  &  0.2534(7)  & 2.01(14)    &  1.16(9)              & 0.76/0.55\\
     \hline
               28 &   0.253(2)  & 2.1(3)    &  1.2(2)              & 0.86/0.46\\
     \hline
    \hline
     \end{tabular}
     \label{tab:fitlog_mod}
 \end{table}


\textit{Conclusion.}---
In this paper, we have studied the scaling behavior of the R\'enyi EE $S_2(L)$ with smooth 
boundaries  at the phase transition point of the 2D $J-Q_3$ model.
We have shown the presence of a logarithmic correction to the area law.
Using the inertia moment density $I(L)$ and spin stiffness $\rho_{s}(L)$ as inputs, 
we have found that the number of Goldstone modes, which is related to the coefficient of the logarithmic term, is four 
by fitting our modified scaling formula to the $S_2(L)$. 
These results indicate
the existence of an emergent $SO(5)$ symmetry at the transition point, which spontaneously breaks to $O(4)$
in the thermodynamic limit but restored in a finite size. 
This result demonstrates that the believed DQCP of the $J-Q_{3}$ model is a 
weak first-order transition point. 
We have also found that the transition point of the $J-Q_2$ model has similar properties \cite{dengjq2},
suggesting the VBS-N\'eel transition in this model is also weakly first order.

With this work, we have provided a new way to distinguish an ordered phase with continuous symmetry broken 
from a critical phase,
which is potentially useful to identify other weak first-order phase transitions, which are hard to determine 
using conventional methods.


This work was supported by the National Natural Science Foundation of China under Grant No.~12175015, No.~12304171, No.~12088101 and 
	MOST 2022YFA1402701 and the Beijing Institute of Technology Research Fund Program for Young Scholars.
The authors acknowledge the support of Tianhe 2JK at the Beijing Computational Science Research Center(CSRC) and the Super Computing Center of Beijing Normal University.
\bibliography{ref}

\clearpage
\section{Supplemental Material}

\subsection{Additional scaling analysis at $Q/J=1.49153$}
\label{fssqc}
Here, we present additional fitting results of Eq. (\ref{newS2}) to $S_2(L)$ at $Q/J=1.49153$.   In the main text, 
$I(L, S=1)$ is used as known finite-size inertia moment density $I(L)$. Table \ref{tab:fitlog_mod2} and \ref{tab:fitlog_mod3} present 
results of fitting to $S_2(L)$ using $I(L, S=2)$ and $I(L, S=3)$ as $I(L)$, respectively. 
The fits remain stable upon further excluding small size points by increasing $L_{min}$ gradually,
although error bars on the fit parameters increase rapidly.
We obtain $N_G/2 =2.03(14)$ and $2.01(14)$, in the best estimations.
We find that our modified formula works for $I(L, S=2)$ and $I(L, S=3)$, which give the same results as that of $I(L, S=1)$.

 \begin{table}[thb]
  \caption{ 
   Fitting results of Eq. (\ref{newS2}) to $S_2(L)$ at $Q/J=1.49153$. In the fits, finite-size $\rho_s(L)$ and $I(L,S=2)$ are used as inputs.  The 
	 range of the system sizes is from $L=8$ to $L=48$.}
    \begin{tabular}{c|c|c|c|c}
    \hline
     \hline
         $L_{min}$ &  $a$    & $b$& $\gamma_{ord}$&$\chi_r^2$/P-value \\   
    \hline
                12&  0.2533(3)  &  2.00(4) &   1.15(2)                & 1.33/0.23  \\
    \hline
                16 & 0.2539(4)   & 1.89(6)   & 1.09(3)                 &0.60/0.73\\
   \hline
               20&  0.2538(5)   & 1.91(10)   &  1.10(5)                & 0.71/0.61 \\     
    \hline
     24 &   0.2532(7)  & 2.03(14)    &  1.16(7)              & 0.54/0.70\\
     \hline
     28 &   0.2526(11)  & 2.16(24)    &  1.22(12)              & 0.58/0.63\\
     \hline
    \hline
     \end{tabular}
     \label{tab:fitlog_mod2}
 \end{table}
 
 \begin{table}[thb]
  \caption{ 
  Fitting results of Eq. (\ref{newS2}) to $S_2(L)$ at $Q/J=1.49153$. In the fits, finite-size $\rho_s(L)$ and $I(L, S=3)$ are used as inputs.  The 
	 range of the system sizes is from $L=8$ to $L=48$.}
    \begin{tabular}{c|c|c|c|c}
    \hline
     \hline
         $L_{min}$ &  $a$    & $b$& $\gamma_{ord}$&$\chi_r^2$/P-value \\   
    \hline
                12&  0.2531(3)  &  2.02(4) &   1.15(2)                & 1.89/0.07  \\
    \hline
                16 & 0.2538(3)   & 1.88(6)   & 1.07(3)                 &0.54/0.77\\
   \hline
               20&  0.2538(5)   & 1.88(10)   &  1.07(5)                & 0.65/0.66 \\     
    \hline
     24 &   0.2532(7)  & 2.01(14)    &  1.13(12)              & 0.43/0.79\\
     \hline
      28 &   0.2527(11)  & 2.13(24)    &  1.20(12)              & 0.45/0.72\\
     \hline
    \hline
     \end{tabular}
     \label{tab:fitlog_mod3}
 \end{table}

Here, we analyze the finite-size behavior of $I(L, S)$ and $\rho_s(L)$. Table \ref{tableI1} shows
a polynomial fit to $I(L, S=1)$ (shown in Fig. \ref{fig:ILqc}).  We find $I(L, S=1)$ converges to a small but finite value, ten times larger
than the error bars. 
Table \ref{tablerho1} shows a polynomial fit to $\rho_s(L)$ (shown in Fig. \ref{chirhos_H}).  We see $\rho_s(L)$ converges to a finite value.

\begin{table}[thb]
	\caption{ Polynomial fit $I(L)=a+b/L+c/L^2$ to the data of $I(L, S=1)$.}
    \begin{tabular}{c|c|c|c|c}
    \hline
     \hline
         $L_{min}$ &  $a$    & $b$& $c$&$\chi_r^2$/P-value \\   
    \hline
                16 &  0.000461(8)  & 0.1681(4)  &   -0.191(5)                & 1.26/0.27   \\
    \hline
                20& 0.000448(13)   & 0.1690(8)   & -0.203(10)                  &1.17/0.32\\
   \hline
               24 &  0.000417(22)   & 0.1711(14)   &  -0.237(21)                & 0.64/0.64 \\
      \hline
                28& 0.000417(40)   & 0.1711(29)    &   -0.237(50)                & 0.85/0.47\\     
    \hline
        \hline
     \end{tabular}
     \label{tableI1}
 \end{table}

\begin{table}[thb]
	\caption{ Polynomial fit $\rho_{s}(L)=a+b/L+c/L^2$ to the data of $\rho_s(L)$.}
    \begin{tabular}{c|c|c|c|c}
    \hline
     \hline
         $L_{min}$ &  $a$    & $b$& $c$&$\chi_r^2$/P-value \\   
    \hline
                20 &  0.0129(3)  & 2.54(2)  &   -5.0(2)                & 1.63/0.15   \\
    \hline
                24& 0.0124(5)   & 2.57(3)   & -5.5(5)                  &1.68/0.15\\
   \hline
               28 &  0.0114(8)   & 2.64(6)   &  -6.9(10)                & 1.48/0.22 \\
     
    \hline
        \hline
     \end{tabular}
     \label{tablerho1}
 \end{table}

\begin{table}[thb]
	\caption{ Power-law fit $I(L)=a L^{b}$ to the data of $I(L, S=1)$.} 
    \begin{tabular}{c|c|c|c}
    \hline
     \hline
         $L_{min}$ &  $a$    & $b$& $\chi_r^2$/P-value \\   
    \hline
                24 &  0.1174(3)  & -0.881(1)                & 1.26/0.27   \\
    \hline
                28& 0.1166(5)   & -0.879(1)                    &1.60/0.17\\
   \hline
               32 &  0.1155(8)  & -0.877(2)                & 1.07/0.36 \\
       \hline
        \hline
     \end{tabular}
     \label{tableI2}
 \end{table}

However, power-law fitting to $I(L, S)$ and $\rho_s(L)$ are also possible for sufficiently large 
$L_{min}=24$ and $28$, respectively. 
For sufficient large $L_{min}$, the fits are also statistically sound. 
The results are listed in Tab.~\ref{tableI2} and Tab.~\ref{tablerho2} 

For conventional $(2+1)$D critical behavior, $\rho_s(L)$ should scale as $1/L$.
It has been found that $L\rho_s$ diverges slowly at the VBS-N\'eel transition point of the $J-Q_2$ model
\cite{Jiang_2008}. One interpretation of this unusual scaling behavior is that the transition is first order.
The scaling is also explained by an unconventional two-length scales scaling scheme\cite{shao_science},
with $\rho_s(L)\sim L^{-\nu/\nu'}\sim L^{-0.715}$, where $\nu$ and $\nu'$ are the correlation length exponent and the exponent 
associated with the divergence of the thickness of VBS domain wall, respectively.
Here we find similar scaling behavior of $\rho_s(L)$ for the $J-Q_3$ model with an exponent close to that in the $J-Q_2$ model. 
However, in this work, we do not try to determine or explain the finite-size scaling behavior of $\rho_s(L)$ and $I(L,S)$. Instead,
we use the finite-size value of $\rho_s(L)$ and $I(L,S)$ as inputs of the fitting formula Eq. (\ref{newS2}).
The results of the fits, in turn, support $I(L)$ and $\rho_s(L)$ have finite values at thermodynamic limit.


\begin{table}[thb]
	\caption{ Power-law fit $\rho_{s}(L)=a L^b$ to the data of $\rho_s(L)$.} 
    \begin{tabular}{c|c|c|c}
    \hline
     \hline
         $L_{min}$ &  $a$    & $b$ &$\chi_r^2$/P-value \\   
    \hline
                28 &  1.339(8)   &   -0.788(2)                & 1.61/0.17   \\
    \hline
                32& 1.322(11)   & -0.784(3)                   &0.84/0.47\\
   \hline
               36&  1.330(18)  & -0.786(4)                   & 1.11/0.33 \\
     
    \hline
        \hline
     \end{tabular}
     \label{tablerho2}
 \end{table}

 
\subsection{ Scaling analyses at $Q/J=1.489$  }
\begin{figure}[h]
    \centering
    \includegraphics[width=0.45\textwidth]{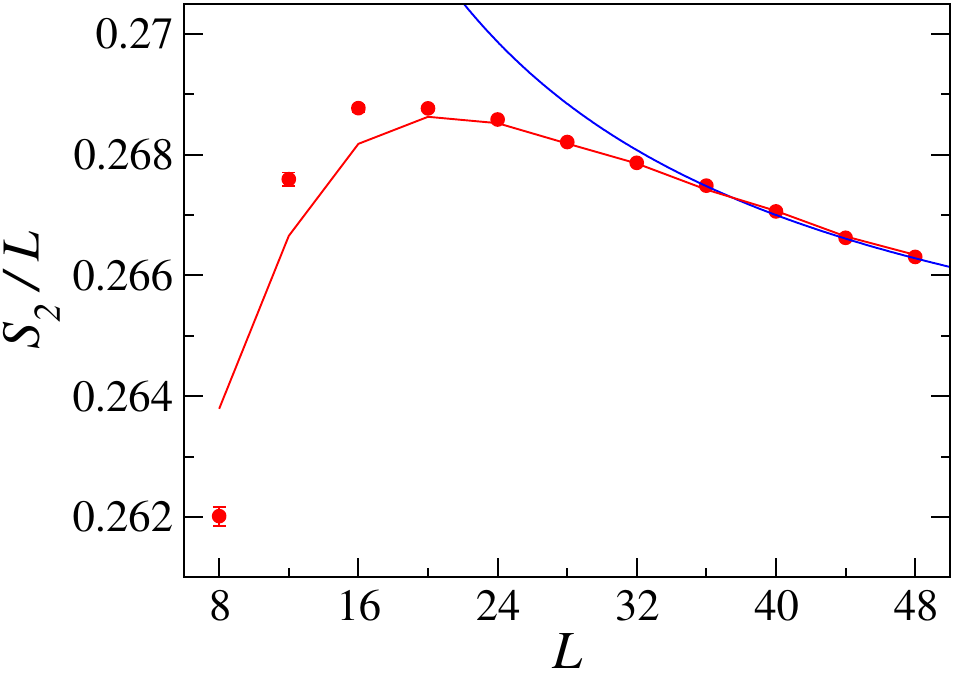}
    \caption{$S_2(L)/L$ vs $L$ at $Q/J=1.489$.  
	The red solid line is the fit using Eq. \ref{newS2} for $L \ge L_{min}=20$, see Tab. \ref{tab:fitlog_mod11}.
	The blue solid line shows a fit without logarithmic correction for $L \ge L_{min}=36$}
    \label{fig:S2-L-1489}
\end{figure}

Here we present results for $S_2(L)$, $\rho_s(L)$, and $I(L,S)$ at $Q/J=1.489$.

Figure \ref{fig:S2-L-1489} shows $S_2(L)/L$ versus system size $L$ at $Q/J=1.489$.   
Table \ref{tab:arealaw1} shows fits to $S_2(L)$ without logarithmic corrections. 
Fits using Eq. (\ref{S2log}) to the same set of data $S_2(L)$ are listed in Tab. \ref{tab:fitlog1}.  
The presence of the logarithmic term is apparent. 

\begin{table}[thb]
	\caption{ Fits without logarithmic correction $S(L)=a L+c$ to the $S_2(L)$ data at $Q/J=1.489$. }
    \begin{tabular}{c|c|c|c}
     \hline
     \hline
        $L_{min}$ &  $a$           &  $c$                      &$\chi_r^2$/P-value \\ 
     \hline
                36&  0.2627(2)     &   0.172(6)                & 0.84/0.43   \\  
     \hline
                40&  0.2625(3)     &   0.182(11)               & 0.40/0.52   \\
     \hline
    \hline
     \end{tabular}
     \label{tab:arealaw1}
 \end{table}

\begin{table}[thb]
	\caption{  Fits using Eq. (\ref{S2log}) to the $S_2(L)$ data. }
    \begin{tabular}{c|c|c|c|c}
    \hline
     \hline
        $L_{min}$ &  $a$         & $b$        & $c$                       &$\chi_r^2$/P-value \\   
    \hline
                8 &  0.2584(1)   & 0.197(3)   &  -0.380(6)                & 2.04/0.037   \\
    \hline
               12 &  0.2584(2)   & 0.195(5)   &  -0.377(11)               & 2.32/0.023\\
    \hline
               16 &  0.2582(3)   & 0.203(7)   &  -0.394(16)               & 2.37/0.028 \\
    \hline
               20 &  0.2573(4)   & 0.231(12)  &  -0.464(28)               & 1.05/0.39\\     
    \hline
               24 &  0.2568(6)   & 0.248(20)  &  -0.508(50)               & 1.02/0.40\\
     \hline
    \hline
     \end{tabular}
     \label{tab:fitlog1}
 \end{table}

Table \ref{tab:fitlog_mod11}, \ref{tab:fitlog_mod12}, and \ref{tab:fitlog_mod13} show fitting results of Eq. (\ref{newS2}) to the $S_2(L)$, using
$I(L, S=1), I(L, S=2)$, and $I(L, S=3)$ as inputs, respectively.
These fits all give $N_G/2=2$ within an error bar of $10 \%$, suggesting the $SO(5)$ symmetry breaking at a first-order transition.

 \begin{table}[thb]
	 \caption{  Fits using Eq. (\ref{newS2}) to the $S_2(L)$ data, in which $I(L, S=1)$ are used as inputs.}
    \begin{tabular}{c|c|c|c|c}
    \hline
     \hline
         $L_{min}$ &  $a$    & $b$& $\gamma_{ord}$&$\chi_r^2$/P-value \\   
    \hline
                12&  0.2549(3)  &  1.74(4) &   1.01(2)                & 2.00/0.05   \\
    \hline
                16 & 0.2551(4)   & 1.71(6)   & 1.00(3)                 &2.23/0.04\\
   \hline
               20&  0.2539(6)   & 1.96(10)   &  1.12(5)                & 0.80/0.55 \\     
    \hline
              24 &   0.2532(9)  & 2.07(17)    &  1.18(9)              & 0.81/0.52\\
     \hline
                 28 &   0.2525(14)  & 2.22(28)    &  1.25(14)              & 0.95/0.42\\
    \hline             
    \hline
     \end{tabular}
     \label{tab:fitlog_mod11}
 \end{table}

 \begin{table}[thb]
	 \caption{  Fits using Eq. (\ref{newS2}) to the $S_2(L)$ data, in which $I(L, S=2)$ are used as inputs.}
    \begin{tabular}{c|c|c|c|c}
    \hline
     \hline
         $L_{min}$ &  $a$    & $b$& $\gamma_{ord}$&$\chi_r^2$/P-value \\   
    \hline
                12&  0.2545(3)  &  1.83(5) &   1.05(3)                & 2.27/0.03   \\
    \hline
                16 & 0.2549(4)   & 1.75(6)   & 1.01(3)                 &1.99/0.06\\
   \hline
               20&  0.2537(6)   & 1.97(10)   &  1.12(5)                & 0.82/0.54 \\     
    \hline
     24 &   0.2531(9)  & 2.09(17)    &  1.18(9)              & 0.85/0.50\\
     \hline
      28 &   0.2527(14)  & 2.19(28)    &  1.23(14)              & 1.06/0.36\\
     \hline
    \hline
     \end{tabular}
     \label{tab:fitlog_mod12}
 \end{table}

  \begin{table}[thb]
	 \caption{  Fits using Eq. (\ref{newS2}) to the $S_2(L)$ data, in which $I(L, S=3)$ are used as inputs.}
    \begin{tabular}{c|c|c|c|c}
    \hline
     \hline
         $L_{min}$ &  $a$    & $b$& $\gamma_{ord}$&$\chi_r^2$/P-value \\   
    \hline
                12&  0.2544(3)  &  1.83(5) &   1.04(3)                & 2.58/0.01   \\
    \hline
                16 & 0.2550(4)   & 1.72(6)   & 0.99(3)                 &1.76/0.1\\
   \hline
               20&  0.2539(6)   & 1.93(10)   &  1.09(5)                & 0.77/0.57 \\     
    \hline
     24 &   0.2533(9)  & 2.03(16)    &  1.14(8)              & 0.80/0.53\\
     \hline
     28 &   0.2529(14)  & 2.12(27)    &  1.18(14)              & 1.01/0.39\\
     \hline
    \hline
     \end{tabular}
     \label{tab:fitlog_mod13}
 \end{table}

We now present finite-size behavior analysis of $I(L, S)$ and $\rho_s(L)$ here.

\begin{figure}
    \centering
    \includegraphics[width=0.45\textwidth]{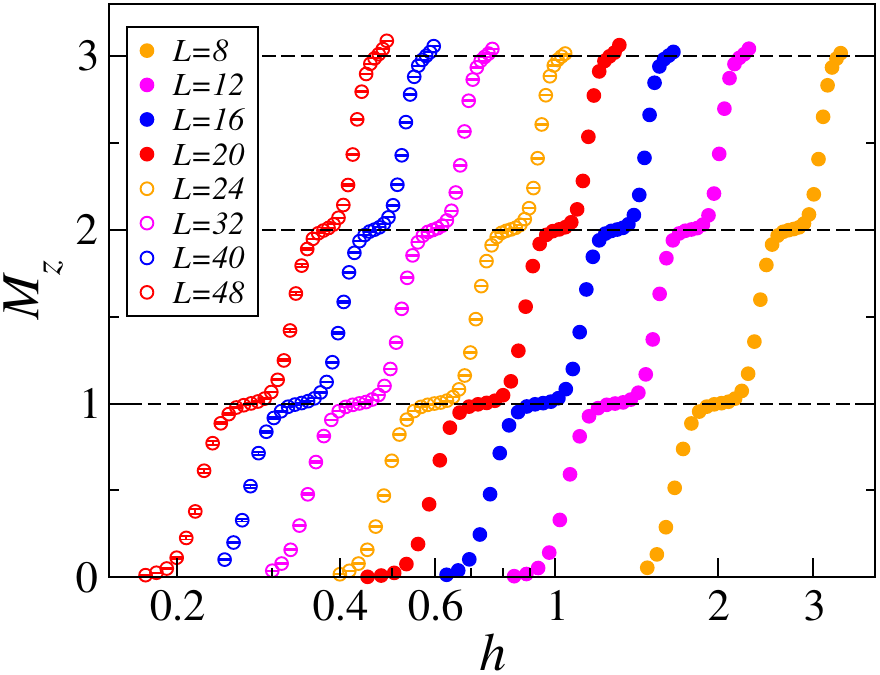}
    \caption{Total magnetization $M_z$ vs external field $h$ at $Q/J=1.489$ for different system size $L$.}
    \label{fig:mzvsh-1489}
\end{figure}

\begin{figure}
    \centering
    \includegraphics[width=0.45\textwidth]{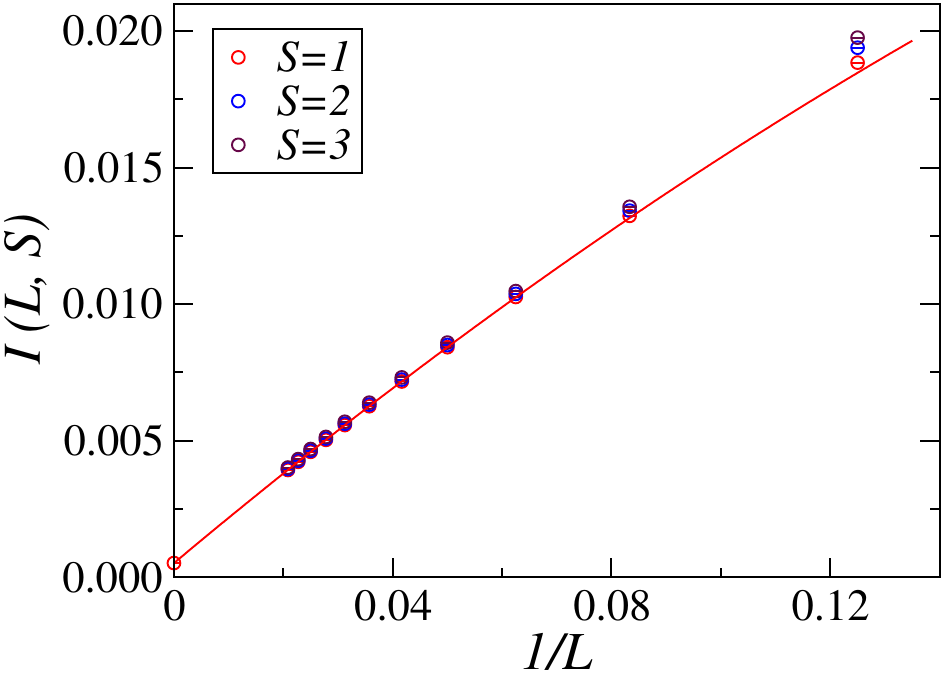}
    \caption{$I(L,S)$ versus $L$ at $Q/J=1.489$. $I(L, S)$ 
	for different $S$ converges to the same value at large $L$, showing the emergence of $SO(5)$ symmetry.
    The red solid line shows a polynomial fit to data points with $L\geq20$ with $I(L=\infty, S=1)$ found to be 0.00052(1), see Tab. \ref{tableI1.489}.}
    \label{fig:IL1489}
\end{figure}

 The total magnetization as a function of the external field at $Q/J=1.489$ for various system sizes is shown in Fig. \ref{fig:mzvsh-1489}.
 We can estimate $I(L, S)$, with $S=1, 2, 3$ from these data. The results are shown in Fig. \ref{fig:IL1489}.

Table \ref{tableI1.489} shows a polynomial fit to $I(L, S=1)$, as shown in Fig. \ref{fig:IL1489}.  Again, we see 
$I(L, S=1)$ converges to a small but finite value, much larger than the error bars. 

\begin{table}[thb]
	\caption{ Polynomial fit $I(L)=a+b/L+c/L^2$ to the data of $I(L, S=1)$.} 
    \begin{tabular}{c|c|c|c|c}
    \hline
     \hline
         $L_{min}$ &  $a$    & $b$& $c$&$\chi_r^2$/P-value \\   
    \hline
                16 &  0.000542(6)  & 0.1663(3)  &   -0.173(4)                & 1.84/0.09   \\
    \hline
                20& 0.000521(10)   & 0.1677(6)   & -0.193(8)                  &0.82/0.54\\
   \hline
               24 &  0.000505(15)   & 0.1688(10)   &  -0.212(15)                & 0.46/0.77 \\
      \hline
                28& 0.000516(26)   & 0.1680(18)    &   -0.197(32)                & 0.51/0.67\\     
    \hline
        \hline
     \end{tabular}
     \label{tableI1.489}
 \end{table}

\begin{figure}[h]
     \centering
     \includegraphics[width=0.45\textwidth]{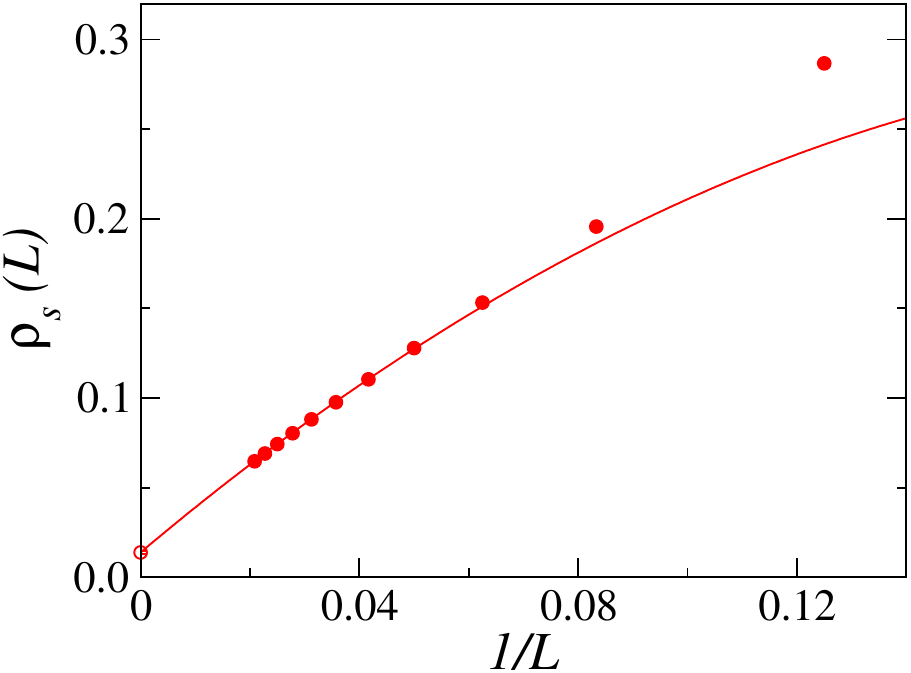}
	 \caption{ Spin stiffness $\rho_s(L)$ versus $1/L$ at $Q/J=1.489$. The red line is a polynomial fit for data points with $L\ge 28$.}
     \label{chirhos_H-1489}
 \end{figure}

 \begin{table}[thb]
	\caption{ Power-law fit $I(L)=a L^{b}$ to the data of $I(L, S=1)$.} 
    \begin{tabular}{c|c|c|c}
    \hline
     \hline
         $L_{min}$ &  $a$    & $b$& $\chi_r^2$/P-value \\   
    \hline
                36 &  0.1093(7)  & -0.859(2)                & 3.10/0.045   \\
    \hline
                40& 0.1071(14)   & -0.854(3)                    &2.96/0.09\\
   \hline              
        \hline
     \end{tabular}
     \label{tableI2.489}
 \end{table}
The calculated spin stiffness $\rho_s(L)$ at $Q/J=1.489$ are shown in Fig. \ref{chirhos_H-1489}.

Table \ref{tablerho1.489} shows a polynomial fit to $\rho_s(L)$.  
We see $\rho_s(L)$ converges to a finite value.

Similar to $Q/J=1.49153$, power-law fitting to $I(L, S=1)$ and $\rho_s(L)$ is also possible for sufficient large $L_{min}$,
although these $L_{min}$ are much larger than those used for polynomial fitting.
The fit results are listed in Tab.\ref{tableI2.489} and \ref{tablerho2.489}, respectively. 
The slow diverging of $L\rho_s$ is seen. 
Again, this behavior is at odds with the conventional $(2+1)$D critical behavior: $\rho_s(L) \propto 1/L$.
Different interpretations are present.  
Still, We do not try to determine or explain the finite-size scaling behavior of $\rho_s(L)$. Instead,
we use the finite-size value of $\rho_s(L)$ and $I(L, S)$ as inputs of the fitting formula Eq. (\ref{newS2}).
The results of the fits, in turn, support $I(L)$ and $\rho_s(L)$ have finite values at the thermodynamic limit.

\begin{table}[thb]
	\caption{ Polynomial fit $\rho_{s}(L)=a+b/L+c/L^2$ to the data of $\rho_s(L)$.}
    \begin{tabular}{c|c|c|c|c}
    \hline
     \hline
         $L_{min}$ &  $a$    & $b$& $c$&$\chi_r^2$/P-value \\   
    \hline
                24 &  0.0153(5)  & 2.46(3)  &   -4.4(5)                & 2.47/0.04   \\
    \hline
                28& 0.0139(8)   & 2.57(6)   & -6(1)                  &1.60/0.19\\
   \hline
               32&  0.0151(14)   & 2.48(11)   &  -4(2)                & 1.91/0.15 \\
    \hline
        \hline
     \end{tabular}
     \label{tablerho1.489}
 \end{table}

\begin{table}[thb]
	\caption{ Power-law fit $\rho_{s}(L)=a L^b$ to the data of $\rho_s(L)$.} 
    \begin{tabular}{c|c|c|c}
    \hline
     \hline
         $L_{min}$ &  $a$    & $b$ &$\chi_r^2$/P-value \\   
    \hline
                36 &  1.20(2)   &   -0.754(4)                & 1.54/0.22   \\
    \hline
                40& 1.20(3)   & -0.755(6)                   &3.04/0.08\\
   \hline     
        \hline
     \end{tabular}
     \label{tablerho2.489}
 \end{table}
\clearpage

\end{document}